\documentclass[]{svjour3}  
\usepackage{epsfig} 
\usepackage{graphicx}
\usepackage{dcolumn}
\usepackage{bm}
\usepackage[]{color}
\usepackage{ulem} 
\usepackage{amsmath}
\usepackage{amssymb}
\usepackage[]{natbib}
\usepackage{float}

\journalname{Bulletin of Mathematical Biology}

\begin{document}

\title{Analysis and control of pre-extinction dynamics in stochastic populations}

\author{Garrett Nieddu  \and Lora Billings \and Eric Forgoston} 

\institute{Garrett Nieddu \at
      Department of Earth and Environmental Sciences,
      Montclair State University,
      1 Normal Avenue,
      Montclair, NJ 07043, USA\\
      \email{nieddug1@mail.montclair.edu}
\and Lora Billings \at
      Department of Mathematical Sciences, 
      Montclair State University,
      1 Normal Avenue,
      Montclair, NJ 07043, USA\\
      \email{billingsl@mail.montclair.edu}
 \and Eric Forgoston \at
      Department of Mathematical Sciences, 
      Montclair State University,
      1 Normal Avenue,
      Montclair, NJ 07043, USA\\
      \email{eric.forgoston@montclair.edu}     
}

\date{Received: date / Accepted: date}

\maketitle

\begin{abstract}
We consider a stochastic population model where the intrinsic or demographic
noise causes cycling between states before the population eventually goes
extinct.  A master equation approach coupled with a WKB
(Wentzel-Kramers-Brillouin) approximation is used to construct the optimal
path to extinction.  In addition, a probabilistic argument is used to
understand the pre-extinction dynamics and approximate the mean time to
extinction.  Analytical results agree well with numerical Monte Carlo
simulations. A control method is implemented to decrease the mean time to extinction. Analytical results quantify the effectiveness of the control and agree
well with numerical simulations.
\end{abstract}

\section{Introduction} \label{sec:intro}
It has long been known that noise can significantly affect physical and biological dynamical systems at a wide variety of levels.  For example, in biology, noise can play a role in sub-cellular processes, tissue dynamics, and large-scale population dynamics~\citep{Tsimring_2014}. Stochasticity can arise in a number of ways.  For example, in epidemiological models, noise is due to the random interactions of individuals in a population as well as uncertainty in epidemic parameter values~\citep{Rand1991,BillingsBS02}. In population ecology, noise may be the result of environmental factors including climatic effects, natural enemies, or inter-specific competition, or may be due to demography~\citep{CRP2004}. 

Stochasticity manifests itself as either external or internal noise. External
noise comes from a source outside the system being considered (e.g the  growth
of a species under climatic effects), and often is modeled by replacing an
external parameter with a random process. 
Internal noise is inherent in the system itself and is caused by the random
interactions of discrete particles
(e.g. individuals in a population). In this article, we are interested in the
dynamics of an isolated single-species population undergoing a set of random
gain-loss processes that simulate births and deaths. Thus, in this
  particular case, the internal noise of the population model is demographic noise.  Mathematically, the effects of these
random interactions are described using a master
equation~\citep{vanKampen_book}. Small fluctuations captured in this modeling
approach may act as an effective force that drives the population to
extinction~\citep{Assaf2010}. While population extinction is assumed to be a
rare event, we can study these models to theoretically understand
pre-extinction dynamics and extinction risk.

\begin{figure}
\begin{center}
\includegraphics[scale=0.4]{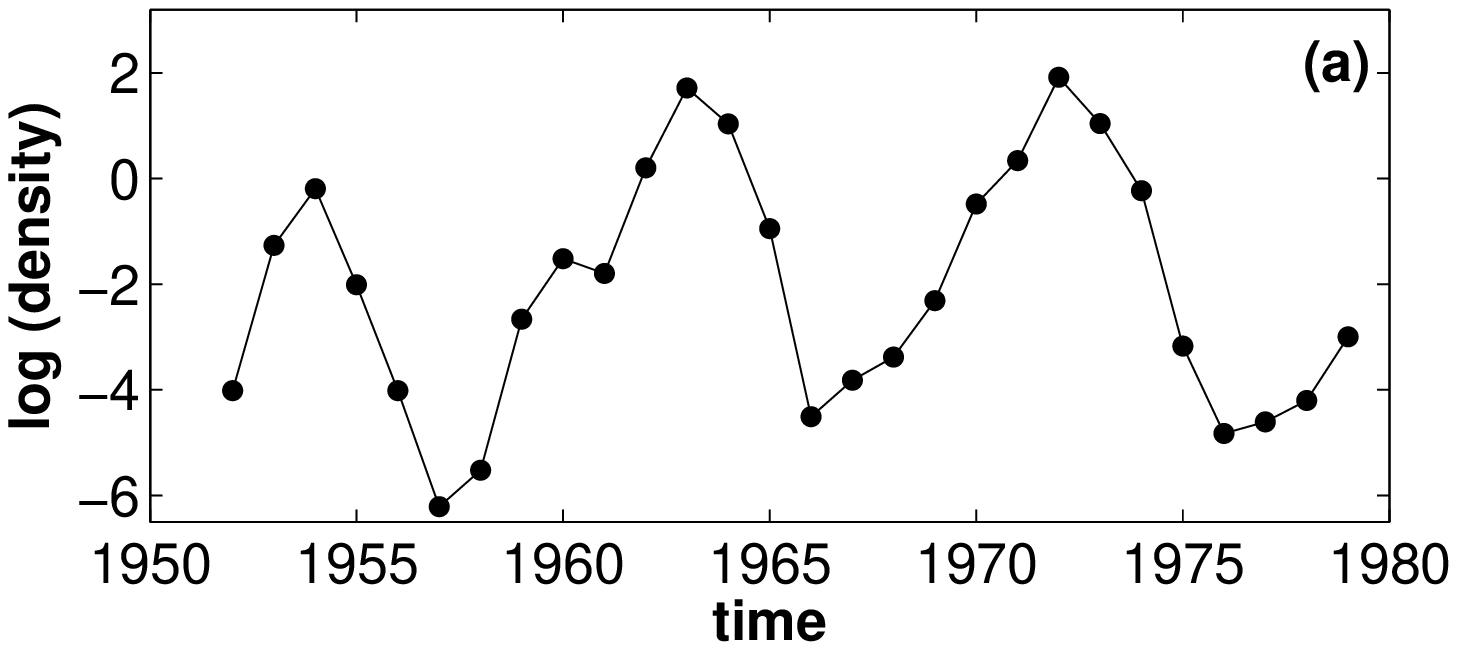}
\includegraphics[scale=0.4]{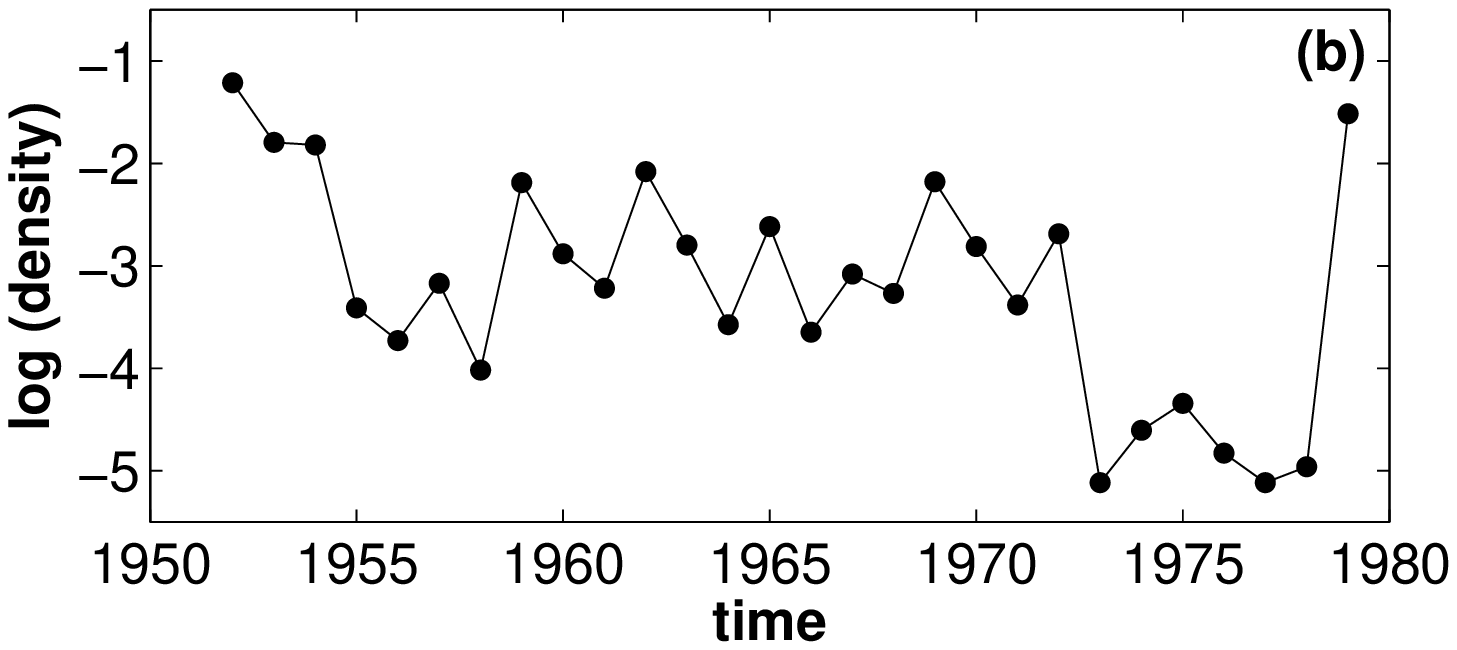}
\caption{\label{fig:moth} Population density fluctuations of Lepidoptera  feeding on larch foliage in the Oberengadin Valley of Switzerland. Data from~\citep{baltensweiler1991_moth} reported as the natural logarithm of numbers per 1000 kg of larch foliage:
(a) Exapate duratella (Tortricidae) and (b) Teleia saltuum (Gelechidae).}
\end{center}
\end{figure}

Extinction risk is an important question in both population dynamics~\citep{Thomas2004} and
ecological community dynamics~\citep{Ebenman2005}. For single populations, it has long
been recognized that extinction risk is dependent on the carrying capacity of
the model~\citep{leigh1981MTE,lande1993MTE}. Mathematically, the carrying
capacity is a positive stable equilibrium in the deterministic model. In this
paper, we consider a population with multiple deterministic stable equilibria
where the population size can stochastically fluctuate between these
states. One such example can be found in the data of forest
Lepidoptera~\citep{baltensweiler1991_moth} 
as shown in
Fig.~\ref{fig:moth}. The drastic population shifts are attributed to a mix of
parasitoids, viral outbreak among the moths, and the quality of the available
foliage~\citep{Bman_moth}. Figure~\ref{fig:moth}(a) shows that the fluctuations may be
regular, although they are not seasonal, while Fig.~\ref{fig:moth}(b) shows a switch between
two states, with vastly different residence times in each state. Similar fluctuations are observed in the context of
human physiology. For example, neural switching and enzyme level cycling are
explored
in~\citep{elf2004bistable,berndt2009bistable,samoilov2005bistable}. Regardless
of the specific context, it is useful to increase our understanding of pre-extinction
dynamics, as well as the mean time to extinction and the path that optimizes
the probability of extinction.

In this article, we explore all of these features for a stochastic model that
exhibits extinction and for a stochastic model that exhibits pre-extinction
cycling that serves to delay the extinction event.
The layout for the article is as follows. Section \ref{gen} presents the
master equation formalism needed to investigate demographic noise in
population models and the general method to find the mean time to
extinction. A simple population model exhibiting extinction is presented in
Section \ref{simple_ext}. It provides an example of the analytical and
numerical methods used to find the mean time to extinction. Pre-extinction
cycling is considered in Section \ref{sec:cycling} and the probabilistic
result used to describe pre-extinction dynamics is derived. A control term is introduced in Section \ref{sec:control} to increase the rate of extinction in the model. The methods from the previous sections are used to quantify the effects of the control. In the last section, we generalize these results and give a brief discussion.

\section{General Theory}\label{gen}
\subsection{Master Equation Formalism}

As mentioned in the introduction, to study the effects of internal noise on the dynamics of a population, a stochastic model must be considered. If the transitions between states are short and uncorrelated, the system is a Markov process and the evolution of the probability is described by a master equation. In the master equation formulation, the probability of the system taking on a particular state $X$ (number of agents), at a given time $t$, is described by $\rho(X,t)$. Let $W(X;r)$ represent the transition rate from a state $X$ to $X+r$, where $r$ can be a positive or negative integer. In this case the time evolution of $\rho(X,t)$ can be written as~\citep{vanKampen_book}:
\begin{equation}
\label{me}
\frac{d \rho(X,t)}{d t}=\sum_{r}\left[ W(X-r;r)\rho\left(X-r,t\right) - W(X;r)\rho(X,t) \right].
\end{equation}
We introduce a rescaled coordinate $x=X/K$, where $K$ is the large parameter
of the problem.  The transition rates are represented as the following expansion in $K$:
\begin{equation}
W(X;r)\equiv W(Kx;r)= Kw_r(x)+u_r(x)+\mathcal{O}(1/K),
\label{trans-rates}
\end{equation}
where $x=\mathcal{O}(1)$, and $w_r(x)$ and $u_r(x)$ also are $\mathcal{O}(1)$.

For $K\gg 1$ the WKB (Wentzel-Kramers-Brillouin) approximation for the scaled
master equation 
can be used~\citep{kubo1973,gang87,Dykman1994d,elgkam04,Kessler07,fbss11,sfbs11}.
Accordingly, we look for the probability distribution in the form of the WKB ansatz
\begin{equation}
\label{eikonal}
 \rho(x,t) = \exp (-K {\cal S}(x,t)),
\end{equation}
where ${\cal S}(x,t)$ is a function known as the action. 

We substitute Eq.~\eqref{eikonal} into 
the scaled master equation which contains terms with the form $w_r(x-r/K)$ and
${\cal S}(x-r/K,t)$, where $r/K$ is small. 
By performing a Taylor series expansion of these functions of $x-r/K$, one arrives at the leading order Hamilton-Jacobi equation
$\mathcal{H}(x,p)=0$, where
\begin{equation}
\label{hammy}
\mathcal{H}(x,p)= \sum_{r} w_r(x) \left( \exp (pr) - 1 \right)
\end{equation}
is the effective Hamiltonian, where $p$ is the conjugate momentum and is defined as $p=d{\cal S}/dx$. In this article, we are interested in the special case of a single step process, for which the only values of $r$ are $+1$ and $-1$. The Hamiltonian for a  single step process will have the general form
\begin{equation} \label{gen_Hsingle}
\mathcal{H}(x,p)=  w_{1}(x) \left( \exp (p) - 1 \right) + w_{-1}(x) \left( \exp (-p) - 1 \right) .
\end{equation}

From the Hamiltonian in Eq.~\eqref{hammy}, one can easily derive Hamilton's equations 
\begin{equation}
\dot{x}=\frac{\partial \mathcal{H}(x,p)}{\partial p} , \qquad
\dot{p}=-\frac{\partial \mathcal{H}(x,p)}{\partial x}.
\label{hes}
\end{equation}
The $x$ dynamics along the $p=0$ deterministic line can be described by the equation
\begin{equation}
\dot{x}=\left . \frac{\partial \mathcal{H}(x,p)}{\partial p}\right
|_{p=0}=\sum\limits_r rw_r(x),
\label{ham-p0}
\end{equation}
which is simply the rescaled mean-field rate equation associated with the
deterministic problem.  For a  single step process, this simplifies to
$ \dot{x}=  w_{1}(x) - w_{-1}(x)$.

\subsection{Mean Time to Extinction}
\begin{figure}
\begin{center}
\includegraphics[scale=0.6]{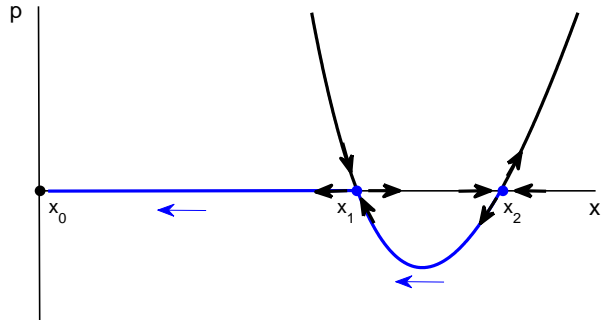}
\caption{\label{fig:Allee3_optpath}Zero-energy trajectories $p=0$, $x=0$, and
  $p_{opt}(x)$ of the Hamiltonian for the stochastic Allee population model
  given by Eq.~\eqref{ham3}.  The optimal path to extinction (blue curve) consists of the
  heteroclinic trajectory $p_{opt}(x)$ (Eq.~\eqref{Allee3_popt})  connecting
  $x_2$ to $x_1$, and the $p=0$ line from $x_1$ to the extinct state
  $x_0$.  }
\end{center}
\end{figure}

We are interested in how intrinsic noise can cause extinction
events of long-lived stochastic populations.  In this article, the extinct
state $x_0$ is an attracting point of the deterministic mean-field equation.
Furthermore, there is an intermediate repelling point $x=x_1$ between the
attracting extinct state and another attracting point $x=x_2$.  This scenario
can be visualized in Fig.~\ref{fig:Allee3_optpath} and corresponds to a scenario
B extinction as explored in~\citep{Assaf2010}.

In this extinction scenario, the most probable path to extinction, or optimal
path to extinction, is composed of two segments.  The first segment is a
heteroclinic trajectory with non-zero momentum that connects the equilibrium
point $(x,p)=(x_2,0)$, where $x_2$ is an attracting fixed point of the
deterministic mean-field equation, with an intermediate equilibrium point
$(x,p)=(x_1,0)$, where $x_1$ is a repelling fixed point of
the deterministic mean-field equation.  The second segment consists of the
segment along $p=0$ from $x_1$ to the extinct state $x_0$.  
 
The optimal path to extinction $p_{opt} (x)$ between $(x_2,0)$ and $(x_1,0)$ is a zero
energy phase trajectory of the Hamiltonian given by Eq.~\eqref{hammy}. 
In a single step process, the optimal path will always have the general form
\begin{equation} \label{gen_popt}
p_{opt}(x)=\ln { \left( w_{-1}(x) /w_1(x)  \right)}.
\end{equation}
Using the definition of the conjugate momentum $p=d{\cal S}/dx$, the action ${\cal S}_{opt}$
along the optimal path $p_{opt} (x)$ is given by
\begin{equation} 
\label{Sopt}
{\cal S}_{opt} = \int_{x_2}^{x_1} p_{opt} (x) dx . 
\end{equation} 
Therefore, the mean time to extinction (MTE) to escape from $(x_2,0)$ and
arrive at $(x_1,0)$  can be approximated by  
\begin{equation}\label{eq:tau_ext_gen} 
\tau = B \exp({K {\cal S}_{opt}}) , 
\end{equation} 
where $B$ is a prefactor that depends on the system parameters and on the
population size. An accurate approximation of the MTE depends on obtaining $B$. 

To capture the deterministic contribution from $x_1$ to $x_0$ in the MTE
approximation, we include the prefactor derived
in~\citep{Assaf2010}. Specifically, the following equation is the general form of the MTE for a single-step scenario B extinction from $x_2$ to $x_0$:
 \begin{equation}\label{eq:tau_ext20} 
\tau_{20} = \frac{2 \pi \exp \left(\int_{x_1}^{x_2}\left(\frac{u_1(x)}{w_1(x)}-\frac{u_{-1}(x)}{w_{-1}(x)} \right)dx\right)}{w_1(x_2)\sqrt{|p'_{opt}(x_1)|p'_{opt}(x_2)}} \exp  \left(K  \int_{x_2}^{x_1} \ln \left( \frac{w_{-1}(x)}{w_1(x)} \right) dx \right) .
\end{equation} 
Note that we will use the general notation $\tau_{ij}$ to identify the
function $\tau_{ij}(x_i,x_{(i+j)/2})$ that provides the escape time from state
$x_i$ to state $x_j$.  In the case that $i=2$ and $j=0$, then one recovers Eq.~\eqref{eq:tau_ext20}.
It is worth noting that the derivation of Eq.~\eqref{eq:tau_ext20} involves
  matching the solution from $x_2$ to $x_1$ asymptotically with the deterministic solution
  from $x_1$ to $x_0$. Because this latter solution is associated with $p=0$,
  its final form does not involve an integral from $x_1$ to
  $x_0$. Nevertheless, the deterministic contribution is in fact included in Eq.~\eqref{eq:tau_ext20}.

\section{An example of extinction}  \label{simple_ext}

To illustrate the analytical methods described in Sec.~\ref{gen}, we consider an example where the local dynamics of a population exhibit the Allee effect.  The Allee effect is seen in animal populations that benefit
from conspecific cooperation. These populations tend to perform better in
larger numbers.  In fact, there is evidence that larger populations are more
capable of avoiding predation, can reproduce faster, and are better able to
resist toxic environmental conditions~\citep{Allee1931,Lidicker2010}. On the
other hand, the growth rate is negative for low densities.  Therefore the
dynamics are bistable and the population will tend towards a positive state, referred to as the carrying capacity, or an extinct state depending on the initial population. A simple deterministic mathematical model demonstrating the Allee effect can be written as
$\dot{x}=f(x)$, where $f(x)$ is a cubic polynomial.  Using the notation used
in Fig.~\ref{fig:Allee3_optpath}, we could write $f(x)=x(x-x_1)(x-x_2)$.
The corresponding stochastic population model is represented by the following
transition processes and associated rates $W(X;r)$.
\begin{center}
\begin{tabular}{ll}
Transition & \hspace{1cm} W(X;r) \\
&\\
${\displaystyle  X \overset{\mu}{\longrightarrow}\varnothing}$  &
\hspace{1cm} $\mu X$,\\
${\displaystyle 2X \overset{\lambda/K}{\longrightarrow}3X}$  & \hspace{1cm} $\lambda \frac{X(X-1)}{2K}$,\\
${\displaystyle 3X \overset{\sigma/K^2}{\longrightarrow}2X}$  &  \hspace{1cm}
$\sigma \frac{X(X-1)(X-2)}{6K^2}$.
\end{tabular}
\end{center}
The first two transitions are required to capture the Allee effect. The death
rate of a low-density population is given by $\mu$, and the growth rate of the
population when the density is large enough is given by $\lambda$. The
negative growth rate for an overcrowded population is provided by $\sigma$,
and $K$ is the carrying capacity of the population.

As described in Sec.~\ref{gen}, the transition processes and their associated
rates are used to formulate the master equation given by Eq.~\eqref{me}.  In
this particular example, all of the transitions are single-step transitions.
Therefore, the increment $r$ only takes on the values of $\pm 1$.  The scaled
transition rates in Eq.~\eqref{trans-rates} are given as
\begin{equation} 
\begin{array}{lll}
w_1(x) = \frac{\lambda x^2}{2}, &~~~~~&
w_{-1}(x) =   \mu x + \frac{\sigma x^3}{6}, \\
&&\\
u_1(x) = -\frac{\lambda x}{2}, &~~~~~&
u_{-1}(x) =   -\frac{\sigma x^2}{2}.
\end{array}
\label{Allee_wr}
\end{equation}
Substitution of Eq.~\eqref{Allee_wr} into Eq.~\eqref{hammy} leads to the
following Hamiltonian:
\begin{equation}
\label{ham3}
\mathcal{H}(x,p)=\frac{\lambda x^2}{2} (e^{p}-1)+ \left( \mu x + \frac{\sigma x^3}{6} \right) (e^{-p}-1) .
\end{equation}
Taking derivatives of Eq.~\eqref{ham3} with respect to $p$ and $x$ (Eq.~\eqref{hes}) lead to the
following system of Hamilton's equations:
\begin{eqnarray}
\dot{x} &=& \frac{\lambda x^2}{2} e^{p} - \left( \mu x+\frac{\sigma x^3}{6} \right) e^{-p} \label{A3xdot}\\
\dot{p}&=& -\lambda x (e^{p}-1)- \left( \mu +\frac{\sigma x^2}{2} \right) (e^{-p}-1) . \label{A3pdot}
\end{eqnarray}

\begin{figure}
\begin{center}
\includegraphics[scale=0.6]{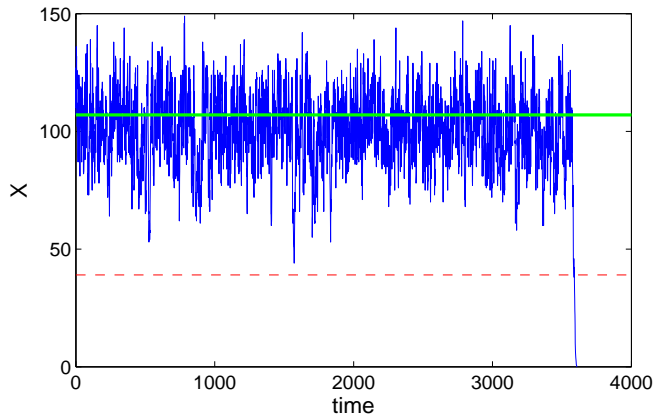}
\caption{\label{fig:ham3_ts}A single realization exhibiting extinction in the
  stochastic Allee population model. The non-zero deterministic stable state is
  shown by the green line, while the deterministic unstable state is
  shown by the red dashed line. The parameter values are $\mu=0.2$, $\sigma=3.0$, $\lambda=1.425$, and $K=100$. }
\end{center}
\end{figure}

By setting the Hamiltonian in Eq.~\eqref{ham3} equal to zero and solving for
$p$ and $x$ it is possible to find three zero-energy phase trajectories.  The
solutions are $x=0$, the extinction line; $p=0$, the
deterministic line and 
\begin{equation}
\label{Allee3_popt}
p_{opt}(x) =\ln{  \left({\frac{6\mu +\sigma x^{2}}{3\lambda x}} \right)},
\end{equation}
the optimal path to extinction. These solutions are shown in Fig.~\ref{fig:Allee3_optpath}. 

Using Eq.~\eqref{ham-p0} we can recover the
deterministic mean-field equation by substituting $p=0$ into Eq.~\eqref{A3xdot} to obtain
\begin{equation}
\label{Allee3_mf} 
\dot{x} = - \frac{\sigma}{6}x^3 + \frac{\lambda}{2}x^2 -\mu x.
\end{equation} 
Equation~\eqref{Allee3_mf} has three steady states: the extinct state $x_0=0$, and two non-zero states
\begin{equation}
\label{Allee_HamEq}
x_{1,2} = \frac {3\lambda\mp\sqrt {9\lambda^2-24\sigma\mu}}{2\sigma}.
\end{equation}
In the deterministic model exhibiting the Allee effect (Eq.~\eqref{Allee3_mf}),
$x_1$ is an unstable steady state that functions as a threshold. For initial
conditions whose value lies between $x_1$ and $x_2$,
the deterministic solution will increase to $x_2$, which is a stable steady
state. For initial conditions whose value is less than $x_1$, the deterministic solution will decrease to the
stable extinct steady state $x_0$. 

However, when intrinsic noise is considered and one performs the analysis
described in Sec.~\ref{gen}, then the steady states of Hamilton's equations
will be two-dimensional with both $p$ and $x$ components.  Furthermore, it is
easy to show that each of the steady states of the stochastic Allee model will
be saddle points, as shown in Fig.~\ref{fig:Allee3_optpath}.
Figure~\ref{fig:Allee3_optpath} shows that starting from $x_2$, the optimal
path to extinction consists of first traveling along the blue heteroclinic
trajectory $p_{opt}(x)$ connecting $x_2$ to $x_1$, followed by traveling along the $p=0$
line from $x_1$ to the extinct state $x_0$. 

\begin{figure}
\begin{center}
\includegraphics[scale=0.7]{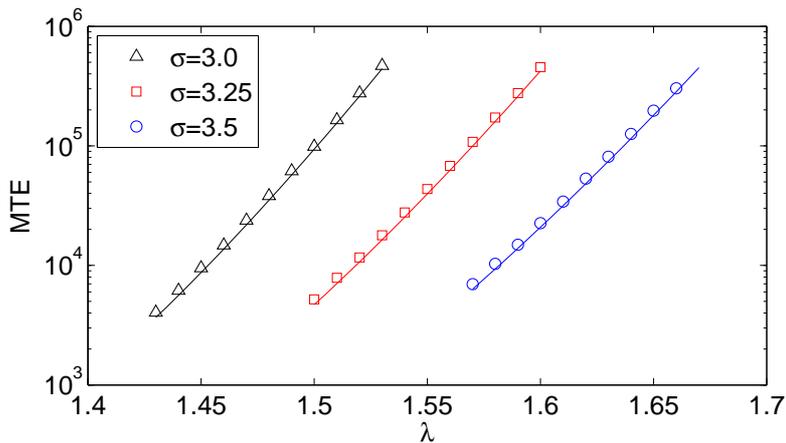}
\caption{\label{fig:ham3_lambda} Mean time to extinction for the stochastic
  Allee population model with an initial population given by $X_2$. The curves
  are found using the analytical approximation given by
  Eq.~\eqref{eq:tau_ext20}, and the symbols represent the corresponding
  numerical simulation results. The numerical results are based on 10,000
  realizations with $\mu=0.2$
  and $K=100$ as $\sigma$ and $\lambda$ are varied.}
\end{center}
\end{figure}

The analytical MTE is found using Eq.~\eqref{eq:tau_ext20} and is confirmed using numerical
simulations.  A Monte Carlo algorithm~\citep{ref:Gillespie76} is used to evolve the population in
time, and Fig.~\ref{fig:ham3_ts} shows an example of one stochastic realization.
Figure~\ref{fig:ham3_ts} shows that the population persists for a very long
time near the $X_2$ state (deterministically stable) but eventually the noise
causes the population to go extinct. By numerically computing thousands of
stochastic realizations and the associated extinction times, one can calculate the MTE. 
 Figure~\ref{fig:ham3_lambda} shows the comparison between the analytical
and the numerical mean time to extinction as a function of $\lambda$ for
various choices of $\sigma$.  Each numerical result is based on
10,000 Monte Carlo simulations, and the agreement is excellent.

 \section{Extinction with Cycling} \label{sec:cycling}

\begin{figure}
\begin{center}
\includegraphics[scale=0.5]{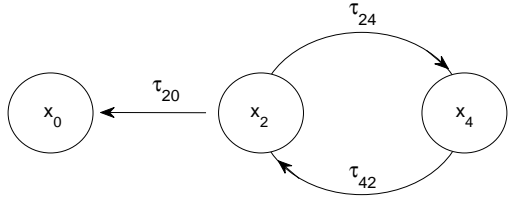} 
\caption{\label{fig:flowchart3}Flowchart for a model whose deterministic mean-field equation has
multiple non-zero stable steady states. The stable states are located at $x_0$ (the extinct state), $x_2$, and $x_4$.  There are two unstable states at $x_1$ and $x_3$ (not shown). The population may stochastically cycle multiple times
from $x_2$ to $x_4$ and back to $x_2$ before eventually transitioning to the
$x_0$ extinct state. }
\end{center}
\end{figure}

In the previous section, we saw that there were three steady states associated
with the deterministic Allee model, two of which were stable (a non-zero carrying
capacity and the extinct state) and one of which was unstable (a threshold
state). Additionally, we saw that for the
stochastic Allee model, the population fluctuated for a long period of time
about one of the deterministically stable states before stochastically switching to the extinct
state. There are many models whose deterministic mean-field equation has
multiple non-zero stable steady states. In these cases, the population can
switch between these different population levels repeatedly before going
extinct. As an example, Fig.~\ref{fig:flowchart3} shows a flowchart with
stable states located at $x_0$ (the extinct state), $x_2$, and $x_4$.  In this
example, there are two unstable states at $x_1$ and $x_3$ (not shown). One can
see from Fig.~\ref{fig:flowchart3} that the population may stochastically cycle multiple times
from $x_2$ to $x_4$ and back to $x_2$ before eventually transitioning to the
$x_0$ extinct state.  In this section, we investigate a population's MTE 
when cycling occurs as a pre-extinction event.  Furthermore, we
derive a new analytical result for the mean time to extinction by considering
the probability of stochastic switching events and their associated transition
times.  This is one of the main results of this article. 

\subsection{An example of population cycling}

It is straightforward to extend the Allee model of Sec.~\ref{simple_ext} to a
model whose deterministic mean-field equation is a quintic polynomial with
five steady states, three of which are stable. The stochastic version of this
new model will exhibit pre-extinction cycling as previously discussed. This stochastic
population model is represented by the following transition processes and
associated rates $W(X;r)$.
\begin{center}
\begin{tabular}{ll}
Transition & \hspace{1cm} W(X;r) \\
&\\
${\displaystyle  X \overset{\mu}{\longrightarrow}\varnothing}$  &
\hspace{1cm} $\mu X$,\\
$2X \overset{\lambda/K}{\longrightarrow}3X$  & \hspace{1cm} $\lambda\frac{X(X-1)}{2K}$,\\
$3X \overset{\sigma/K^2}{\longrightarrow}2X$  &  \hspace{1cm} $\sigma\frac{X(X-1)(X-2)}{6K^2}$, \\
$4X \overset{\alpha/K^3}{\longrightarrow}5X$ &  \hspace{1cm} $\alpha \frac{X(X-1)(X-2)(X-3)}{24K^3}$, \\
$5X \overset{\beta/K^4}{\longrightarrow}4X$ & \hspace{1cm} $\beta\frac{X(X-1)(X-2)(X-3)(X-4)}{120K^4}$.
\end{tabular}
\end{center}
The first three events are the same as found in the stochastic Allee model in
Sec.~\ref{simple_ext} and allow for fluctuations around a population level
before going extinct. The two new events with their associated
positive ($\alpha$) and negative ($\beta$) growth rates allow for fluctuations
around a second population level as well as cycling between the two population
levels before going extinct.

As described in Sec.~\ref{gen}, the transition processes and their associated
rates are used to formulate the master equation given by Eq.~\eqref{me}. Note that 
the transitions are single-step because 
the increment $r$ only takes on the values of $\pm 1$.  Therefore, the scaled
transition rates $w_r(x)$ and $u_r(x)$ in Eq.~\eqref{trans-rates} are given as
\begin{equation}
\begin{array}{lll}
w_1(x) = \frac{\lambda x^2}{2}+\frac{\alpha x^4}{24}, &~~~~~~&
w_{-1}(x) =    \mu x+\frac{\sigma x^3}{6}+{\frac {\beta x^5}{120}}, \\
&&\\
u_1(x) =-  \frac{\lambda x}{2} -\frac{\alpha x^3}{4} , &~~~~~~&
u_{-1}(x) =   - \frac{\sigma x^2}{2}- \frac{\beta x^4}{12}. \\
\end{array}
\label{cycling_wr}
\end{equation}

Substitution of Eq.~\eqref{cycling_wr} into Eq.~\eqref{hammy} leads to the
following Hamiltonian:
\begin{equation}
\label{ham5}
\mathcal{H}(x,p)=  \left( \frac{\lambda x^2}{2}+\frac{\alpha
    x^4}{24} \right)  \left( {e^{p}}-1 \right) + \left( \mu x+\frac{\sigma x^3}{6}+{\frac {\beta x^5}{120}}
\right)  \left( {e^{-p}}-1 \right)  .
\end{equation}
Taking derivatives of Eq.~\eqref{ham5} with respect to $p$ and $x$ (Eq.~\eqref{hes}) lead to the
following system of Hamilton's equations:
\begin{eqnarray}
\dot{x} &=& \left( \frac{\lambda x^2}{2}+\frac{\alpha x
^4}{24} \right) {e^{p}} - \left( \mu x+\frac{\sigma x^3}{6}+{\frac {\beta x^5}{120}}
 \right) {e^{-p}}
, \label{A5xdot}\\
\dot{p}&=&- \left( \lambda x+\frac{\alpha x^
3}{6} \right)  \left( {e^{p}}-1 \right) - \left( \mu+\frac{\sigma x^2}{2}+\frac{\beta x^4}{24} \right) 
 \left( {e^{-p}}-1 \right) 
 . \label{A5pdot}
\end{eqnarray}

\begin{figure}
\begin{center}
\includegraphics[scale=0.6]{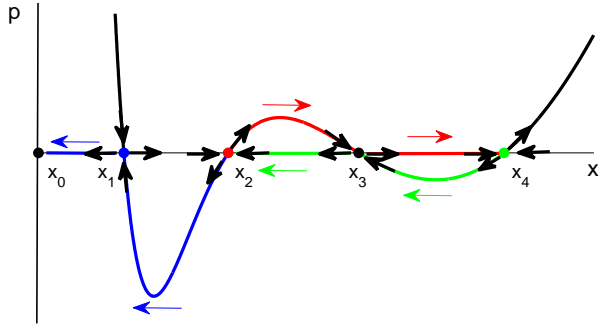}
\caption{\label{fig:Allee5_optpath}Zero-energy trajectories of the Hamiltonian for the stochastic Allee population model
  given by Eq.~\eqref{ham5}.  The optimal path of transitioning from one state
  to another is given by $p=0$ or $p_{opt}(x)$ (Eq.~\eqref{Allee5_popt}). A
  cycling path (red and green) consists of the heteroclinic trajectory  connecting
   $x_2$ to $x_3$ (red) and the $p=0$ line from $x_3$ to $x_4$ (red), followed
   by the heteroclinic trajectory  connecting
   $x_4$ to $x_3$ (green) and the $p=0$ line from $x_3$ to $x_2$ (green).  The
 optimal path to extinction consists of the heteroclinic trajectory from $x_2$ to $x_1$ (blue) and the $p=0$ line from $x_1$ to $x_0$ (blue).}
\end{center}
\end{figure}

Once again, by setting the Hamiltonian in Eq.~\eqref{ham5} equal to zero and
solving for $p$ and $x$ it is possible to find the zero-energy phase
trajectories to be $x=0$, the extinction line; $p=0$, the deterministic line and
\begin{equation} \label{Allee5_popt}
p_{opt}(x)=\ln { \left( {\frac {120\mu+20\sigma x^2+\beta x^4}{
5x \left( \alpha x^2+12\lambda \right) }} \right)},
\end{equation}
the optimal path to extinction.  The $p=0$ and $p_{opt}(x)$ solutions found
using Eq.~\eqref{ham5} are shown in Fig.~\ref{fig:Allee5_optpath}.  Using Eq.~\eqref{ham-p0} we can recover the
deterministic mean-field equation by substituting $p=0$ into Eq.~\eqref{A5xdot} to obtain
\begin{equation}
\label{Allee5_mf} 
\dot{x} = - \frac{\beta}{120}x^5 + \frac{\alpha}{24}x^4-\frac{\sigma}{6}x^3+\frac{\lambda}{2}x^2 -\mu x.
\end{equation} 

Equation~\eqref{Allee5_mf} has five steady states: the extinct state $x_0=0$,
and four non-zero states, two of which are stable and two
of which are unstable.  In the deterministic model given by
Eq.~\eqref{Allee5_mf}, $x_1$ and $x_3$ are unstable states.  For initial
conditions whose value lies between $x_1$ and $x_2$, the
deterministic solution will increase to $x_2$, which is a stable steady state.
For initial conditions whose value is less than $x_1$, the deterministic solution will decrease to the stable
extinct steady state $x_0$.  Similarly, when initial conditions have a value
between $x_3$ and $x_4$, the
deterministic solution will increase to $x_4$, which is a stable steady
state. When initial conditions have a value between
$x_3$ and $x_2$, the deterministic solution will decrease to the stable
steady state $x_2$.  

As we have already seen in the previous section, the inclusion of intrinsic
noise in the model leads to the steady states of Hamilton's equations being
two-dimensional with both $p$ and $x$ components.  Furthermore, the steady
states of the stochastic cycling model will be saddle points, as seen in
Fig.~\ref{fig:Allee5_optpath}.  Figure~\ref{fig:Allee5_optpath} shows that
starting from $x_2$ there is a choice to be made: 1) the population could go
extinct by traveling along the blue path, which is the heteroclinic trajectory connecting $x_2$ to
$x_1$, followed by traveling along the $p=0$ line from $x_1$ to the extinct
state $x_0$, much like what happens in the stochastic Allee model; or 2) the
population could cycle to $x_4$ and back by traveling along the red path and
then the green path, which includes
two stochastic escapes. First, the population travels along the heteroclinic trajectory connecting $x_2$ to
$x_3$, followed by traveling along the $p=0$ line from $x_3$ to $x_4$.  After
fluctuating for some time about $x_4$, the population returns to $x_2$ by
traveling along the heteroclinic trajectory from $x_4$ to $x_3$, followed by
traveling along the $p=0$ line from $x_3$ to $x_2$.

\begin{figure}
\begin{center}
\includegraphics[scale=0.6]{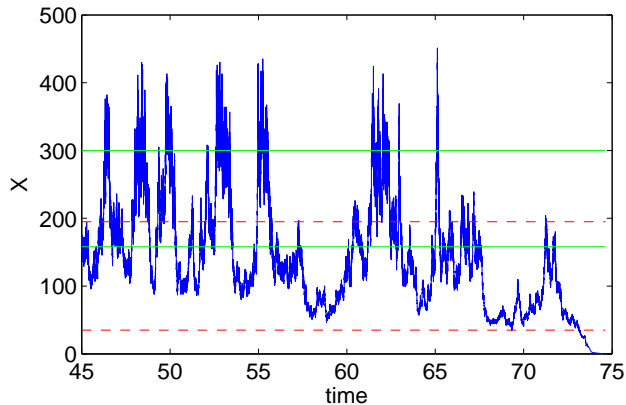}
\caption{\label{fig:ham5_ts} A single realization exhibiting cycling and extinction in the
  stochastic cycling population model. The non-zero deterministic stable states are
  shown by the green lines, while the deterministic unstable states are
  shown by the red dashed lines. The parameter values are $\mu=3.25$, $\alpha=0.465$, $\beta=0.048$, $\lambda=3.96$,
  $\sigma=1.905$, and  $K=14$.}
\end{center}
\end{figure}

A Monte Carlo algorithm~\citep{ref:Gillespie76} is used to evolve the population in
time, and Fig.~\ref{fig:ham5_ts} shows one stochastic realization.
Figure~\ref{fig:ham5_ts} shows multiple cycling events between the $X_2$ and
$X_4$ states (deterministically stable) before the population
eventually goes extinct.  By numerically simulating thousands of
stochastic realizations, we can compute the mean time to extinction and
compare the numerical result with an analytical
result. We will now derive a novel analytical form for the mean time to extinction
that will account for the additional pre-extinction cycling time that delays
the actual extinction event.

\subsection{Approximating the extinction time}

Consider the model whose flowchart is given in Fig.~\ref{fig:flowchart3}.  The
deterministic mean-field equation for this model has multiple non-zero stable
steady states located at $x_2$ and $x_4$ along with the extinct state located
at $x_0$.  In the corresponding stochastic model, the population may
stochastically cycle multiple times from $x_2$ to $x_4$ and back to $x_2$
before eventually transitioning to the extinct state $x_0$.  It is important
to note that in this example it is not possible to experience unlimited
population growth~\citep{MeeSas2008}, and eventually the population will go extinct.

If the system is located at $x_2$, there are only two options for a stochastic
switch: 1) the population can go to the extinct state $x_0$, or 2) the
population can switch to $x_4$. Since the population will eventually go
extinct, it follows that any stochastic switch from $x_2$ to $x_4$ must result
in a following switch from $x_4$ back to $x_2$ at some later
time. Furthermore, the population may cycle from $x_2$ to $x_4$ and back to
$x_2$ any number of times before the population eventually goes extinct by
switching from $x_2$ to the absorbing extinct state $x_0$. 

In isolation, the probability of the population switching from  $x_2$ to $x_0$
can be approximated as $1/\tau_{20}$. Similarly,  the probability of the
population switching from  $x_2$ to $x_4$ can be approximated by
$1/\tau_{24}$. Recall that both $\tau_{20}$ and  $\tau_{24}$ can be
approximated using  Eq.~\eqref{eq:tau_ext20}. However, in the cycling model,
these switches do not occur in isolation.  Rather, there is a ``competition''
as to which switch will happen first.  Therefore, we must compute the
probability of one switch occurring before the other. The probability of the
population switching from  $x_2$ to $x_0$ before switching from $x_2$ to $x_4$ is
\begin{equation} 
\mathcal{P}_{20} = \frac{\frac{1}{\tau_{20}}}{\frac{1}{\tau_{20}}+\frac{1}{\tau_{24}}} = \frac{\tau_{24}}{\tau_{20}+ \tau_{24}} .\label{eq:P20}
\end{equation}
Note that we will use the general notation $\mathcal{P}_{ij}$ to denote the
probability that an escape from  $x_i$ to $x_j$ happens first. Also, 
 $\mathcal{P}_{24} = 1- \mathcal{P}_{20}$ because there are only two switching options. 

To find the MTE, we use a probabilistic argument whereby the probability
  of a given event (immediate extinction, one cycle followed by extinction,
  two cycles followed by extinction, etc.) is weighted by the approximate time
  of each event. Each transition time is found using Eq.~\eqref{eq:tau_ext20},
  and it should be noted that each probability term is written in terms of
  these approximate transition times (e.g. Eq.~\eqref{eq:P20}). The MTE thus becomes the sum of the expected times for all possible number of cycles to occur and the final escape from $x_2$ to $x_0$:
\begin{subequations}
\begin{align}
{\rm MTE}&=  \tau_{20}\mathcal{P}_{20}+ \sum_{i=0}^{\infty}i(\tau_{24}+\tau_{42})(\mathcal{P}_{24})^i\mathcal{P}_{20} 
\label{mte-prob_gen}\\
&=   \tau_{20}\mathcal{P}_{20}+\frac{(\tau_{24}+\tau_{42})\mathcal{P}_{24}}{\mathcal{P}_{20}} \\
&= \frac{\tau_{20}\tau_{24}}{ \tau_{20}+ \tau_{24}}+ (\tau_{24}+\tau_{42})\frac{\tau_{20}}{\tau_{24}}.
\label{mte-prob}
\end{align}
\end{subequations}

\begin{figure}
\begin{center}
\includegraphics[scale=0.6]{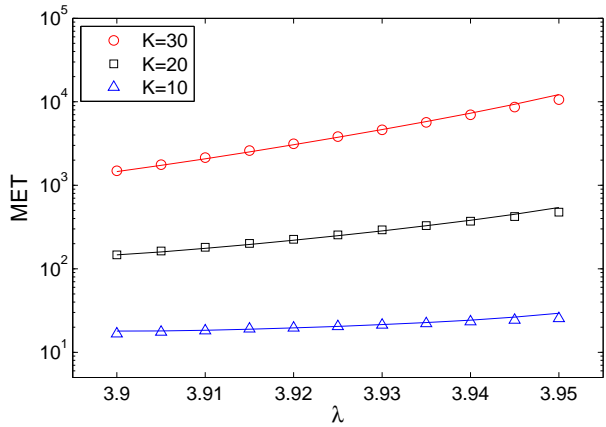}
\caption{\label{fig:ham5_lambda} Mean time to extinction for the stochastic
  cycling population model with an initial population given by $X_2$. The
  solid curves are found using the analytical approximation given by
  Eq.~\eqref{mte-prob}, and the symbols represent the
  corresponding numerical simulation results. The numerical results are based
  on 5,000 simulations with $\mu=3.307$, $\alpha=0.458$, $\beta=0.047$, and
  $\sigma=1.8874$ as $K$ and $\lambda$ are varied. }
\end{center}
\end{figure}

As in the stochastic Allee model, the analytical mean time to extinction for
the stochastic cycling model can be confirmed using Monte Carlo
simulations~\citep{ref:Gillespie76}.  Figure~\ref{fig:ham5_lambda} shows the
comparison between the analytical and numerical mean time to extinction as a
function of $\lambda$ for various choices of carrying capacity $K$.  Each numerical result
is based on 5,000 Monte Carlo simulations, and the agreement is excellent. Note that the choice of $\lambda$ values for this
  example is limited by the quasi-stationarity requirement. However, in the
  following section, we continue the exploration of this example using control
  and show there is excellent agreement for MTE over several orders of
  magnitude of MTE. 

\section{Speeding up Extinction} \label{sec:control}

The previous section presents a way to find the MTE in a population model with
cycling so that extinction is delayed. Often the MTE is of interest in the
study of population dynamics because either longevity or quick extinction has
value. The population studied in this section should be thought of as pests,
and a short MTE should be considered ideal. The control method we model
removes individuals at a particular frequency $\nu$. This population will have
all the same demographic events that were seen in the cycling population model, and will have the following event in addition:
\begin{center}
\begin{tabular}{cc}
Transition & W(X;r) \vspace*{0.05in} \\
${\displaystyle  X \overset{\nu}{\longrightarrow}\varnothing}$  & ${\displaystyle \nu }$.
\end{tabular}
\end{center}
In an ecological context one might think of the control term as culling or quarantining.

The only change from the cycling model transition rates given by Eq.~\eqref{cycling_wr}
is in the rate $w_{-1}$ which now has the form
\begin{equation}
w_{-1}(x) =  \mu x + \frac{\sigma x^3}{6} +\frac{\beta x^5}{120}  + \frac{\nu}{K}.
\end{equation}
Using Eq.~\eqref{hammy}, the modified Hamiltonian will be
\begin{equation}
\label{ham5_control}
\mathcal{H}(x,p)=\left( \frac{\alpha x^4}{24} +  \frac{\lambda x^2}{2} \right)
(e^{p}-1)+ \left( \mu x + \frac{\sigma x^3}{6} +\frac{\beta x^5}{120}  + \frac{\nu}{K}\right) (e^{-p}-1) .
\end{equation}
To quantify the change in the MTE as a function of $\nu$, we use
Eq.~\eqref{mte-prob} and the modified Hamiltonian given by Eq.~\eqref{ham5_control}.

\begin{figure}
\begin{center}
\includegraphics[scale=0.6]{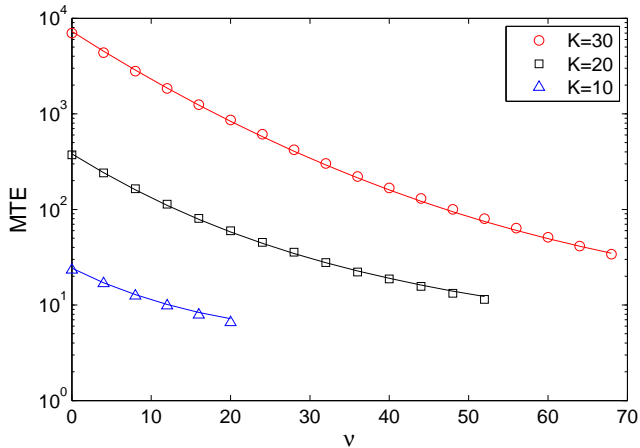}
\caption{\label{fig:ham5_control} Mean time to extinction for the stochastic
  cycling population model using control with an initial population given by $X_2$. The
  solid curves are found using the analytical approximation. The symbols represent the
  corresponding numerical simulation results. The numerical results are based
  on 5,000 simulations with $\mu=3.307$, $\alpha=0.458$, $\beta=0.047$, $\lambda=3.94$, and
  $\sigma=1.8874$ as $K$ and $\nu$ are varied. }
\end{center}
\end{figure}

The analytical mean time to extinction for
the stochastic cycling model with control can be confirmed using Monte Carlo
simulations~\citep{ref:Gillespie76}.  Figure~\ref{fig:ham5_control} extends the examples from Fig.~\ref{fig:ham5_lambda} by comparing the analytical and numerical mean time to extinction as a
function of $\nu$.   Each numerical result
is based on 5,000 Monte Carlo simulations, and the agreement is excellent over several orders of magnitude of MTE. As
expected, when more individuals are removed from the population the MTE decreases. 

\section{Conclusions and Discussion }

In this article, we have considered stochastic population models where the intrinsic or demographic noise eventually causes the population to go extinct. For models that exhibit stochastic cycling between two states, we described the optimal path to extinction and an analytical method to approximate the mean time to extinction. We used a probabilistic argument
to understand the pre-extinction dynamics that delay the extinction event.

These results for the MTE can be extended to the general case of a system with
$2n+1$ steady states (($n \in \mathbb{N}$), $n>1$), with the possibility of
$n-1$ cycles. We assume there are $n+1$ deterministically stable steady states
$\{X_0,X_2,X_4,\ldots,X_{2n}\}$ alternating with deterministically unstable
steady states, and that $X_0$ is an absorbing extinct state. For a system starting at $X_2$, 
\begin{equation} \label{eq:MTE_gen}
MTE =  \tau_{20}\mathcal{P}_{20}+\sum_{i=1}^{n-1}  \left[ \left(\tau_{2i,2i+2}+\tau_{2i+2,2i}\mathcal{P}_{2i+2,2i}\right) \prod_{k=1}^{i} \frac{ \mathcal{P}_{2k,2k+2}}{\mathcal{P}_{2k,2k-2}}  \right].
\end{equation}
Note that when situated at $X_{2n}$, the stable steady state furthest away
from the extinct state, then there is no choice of switching.  The only
possible switch is from $X_{2n}$ to $X_{2n-2}$, and therefore $
\mathcal{P}_{2n,2n-2}=1$. This result can also be extended to find the MTE
when the system starts at any of the other deterministically stable steady states. Consider an initial condition $X_{2k}$ for $k<n$. One would need to find the mean time for each escape in the sequence  $X_{2k},X_{2k-2},\ldots,X_0$. For example, starting at $X_{2k}$ would reduce the problem to the subsystem of deterministically stable steady states $\{X_{2k-2},X_{2k},\ldots,X_{2n}\}$ for which   Eq.~\eqref{eq:MTE_gen} would approximate the mean escape time to $X_{k-2}$. Repeating this procedure leftward and taking the sum of these mean escape times would result in the total MTE. 

Lastly, a control method was introduced to the stochastic cycling population
model. The mean time to extinction was calculated analytically and was shown
to agree well with numerical Monte Carlo simulations. It was shown that the
mean time to extinction decreases monotonically with an increased removal
program. From an ecological perspective it is important to work towards a
quantitative understanding of how control methods (e.g. bio-control agents, culling programs, quarantine programs, or hunting allowances) may affect the longevity of a population.

\begin{acknowledgements}
We gratefully acknowledge support from the National Science Foundation. GN, LB, and EF were supported by the National Science Foundation awards CMMI-1233397 and DMS-0959461. This material is based upon work while LB was serving at the National Science Foundation. Any opinion, findings, and conclusions or recommendations expressed in this material are those of the authors and do not necessarily reflect the views of the National Science Foundation. We also gratefully acknowledge Dirk Vanderklein and Andrew McDougall for helpful discussions.

\end{acknowledgements}


\end{document}